\documentclass[12pt]{article}
\input epsf
\usepackage{epsfig}

\def\Ai{\mathrm{Ai}}
\def\R{\mathbf{R}}
\def\C{\mathbf{C}}

\title{Quasi-exactly solvable quartic: elementary integrals
and asymptotics}
\author{A. Eremenko and A. Gabrielov\thanks{Both authors were 
supported by NSF grant DMS-1067886.}}
\begin{document}
\maketitle
\hfill\parbox{3.5in}{Littlewood, when he makes use of an algebraic identity,
always saves himself the trouble of proving it; he maintains
that an identity, if true, can be verified in few lines by anybody
obtuse enough to feel the need of verification.
\vspace{.1in}

\hfill{\it Freeman Dyson} \cite{D}}
\begin{abstract} We study elementary eigenfunctions
$y=pe^h$ of operators $L(y)=y^{\prime\prime}+Py$,
where $p,h$ and $P$ are polynomials in one variable.
For the case when $h$ is 
an odd cubic polynomial, we investigate the real level crossing points
and asymptotics of eigenvalues.
This study leads to an interesting identity with elementary integrals.


\end{abstract}

\noindent
\section{Introduction}\label{intro}

Following Bender and Boettcher \cite{BB}, we consider the eigenvalue problem
\begin{equation}\label{phys}
w^{\prime\prime}+(\zeta^4+2b\zeta^2+2iJ\zeta+\lambda)w=0,\quad
w(te^{-i(\pi/2\pm\pi/3)})\to 0,\quad t\to+\infty,
\end{equation}
with real parameters $b$ and $J$.

When $J$ is a positive integer, this problem is quasi-exactly solvable (QES)
which means that for every $b$
there are $J$ eigenvalues $\lambda_{J,k},\; k=1,\ldots,J$ 
with {\em elementary} eigenfunctions 
$$y_{J,k}(\zeta)=p_{J,k}(\zeta)\exp(-i\zeta^3/3-ib\zeta),$$
where $p_{J,k}$ are polynomials of degree $J-1$.
These $J$ eigenvalues are found from algebraic equations
\begin{equation}\label{QJ}
Q_J(b,\lambda)=0,\quad J=1,2,\ldots,
\end{equation}
where $Q_J$ are real polynomials of degree $J$ in $\lambda$.

For every $J$ and $b$,
problem (\ref{phys}) has infinitely many eigenvalues $\lambda$
tending to infinity, which satisfy an equation
\begin{equation}\label{FJ}
F_J(b,\lambda)=0,
\end{equation}
where $F_J$ is a real entire function of two variables which is called
the {\em spectral determinant}.

We denote by $Z_J(\R)$ and $Z_J^{QES}(\R)$ the sets of real solutions
$(b,\lambda)$ of equations (\ref{FJ}) and (\ref{QJ}), respectively.
These are certain analytic curves in $\R^2$. A computer-generated picture
of these curves is given in \cite{BB}.
The symbols $Z_J$ and $Z_J^{QES}$ denote the sets of all complex
solutions of these equations.

The most conspicuous feature seen in the picture in \cite{BB}
is a set of real
crossing points where the curves of $Z_J^{QES}(\R)$ intersect
the rest of $Z_J(\R)$.

In this paper, we prove that infinitely many such points exist
for each odd $J$
(Proposition 1). We characterize real and complex
crossing points as intersection
points of $Z_J^{QES}$ with $Z_{-J}$ (Theorem 1).
We prove that all non-QES eigenvalues are real when $b$ is real
and $J$ is a positive integer (Theorem 2.)

In sections \ref{77}, \ref{88}, we study
the asymptotic behavior of polynomials $Q_J$ as $b\to\pm\infty$,
using  singular perturbation methods from \cite{EG2}. Appendix A
with a description of the relevant results of \cite{EG2} is included.

Our proof of Theorem~1 is related to a challenging conjecture
on certain elementary integrals. We rigorously verified this conjecture
for integers $J\leq 5$ using symbolic computation. The conjecture
is discussed in sections \ref{22}, \ref{44}.

\section{Elementary solutions}\label{11}

Let $h$ and $p$ be two polynomials in one variable.
When $y=p(z)e^{h(z)}$ satisfies
a second order differential equation
\begin{equation}\label{1}
y^{\prime\prime}+Py=\lambda y,
\end{equation}
where $P$ is a polynomial?
Substitution gives
\begin{equation}\label{7}
\frac{p^{\prime\prime}}{p}+2\frac{p'}{p}h'+h^{\prime\prime}+{h^\prime}^2+P-\lambda=0.
\end{equation}
Such $P$ exists if and only if 
\begin{equation}\label{cond1}
p^{\prime\prime}+2p'h'\quad\mbox{is divisible
by}\quad p.
\end{equation}
Another criterion is obtained if we consider the second solution $y_1$
of (\ref{1})
which is linearly independent of $y$. This second solution can be found
from the condition
\begin{equation}\label{2}
yy_1^\prime-y'y_1=1.
\end{equation}
Solving (\ref{2}) with respect to $y_1$ we obtain 
\begin{equation}
\label{y1}
y_1=pe^h\int p^{-2}e^{-2h}.
\end{equation}
As all solutions of (\ref{1}) must be entire functions, we conclude
that
\begin{equation}\label{cond2}
\mbox{all residues of}\quad p^{-2}e^{-2h} \quad\mbox{vanish.}
\end{equation}
This condition is {\em necessary and sufficient} for $y=pe^h$ to satisfy
equation (\ref{1}) with some $P$. Indeed, if (\ref{cond2}) holds,
then $y_1$ defined by (\ref{y1})
is an entire function, so $(y,y_1)$ is a pair of entire functions
whose Wronski determinant equals $1$, so this pair must satisfy a differential
equation (\ref{1}) with entire $P$, and asymptotics at infinity show that
$P$ must be a polynomial.

Thus conditions (\ref{cond1}) and (\ref{cond2}) are equivalent.
One can give another equivalent condition in terms of zeros of $p$,
as in \cite{U}.
Let $$p(z)=\prod_{j=1}^n(z-z_j),\quad p_k(z)=p(z)/(z-z_k).$$
Then (\ref{cond1}) is equivalent to
$$p^{\prime\prime}(z_k)+2p'(z_k)h'(z_k)=0,$$
for all $k=1,\ldots,n$. We have $p'(z_k)=p_k(z_k)$ and $p^{\prime\prime}(z_k)=
2p^\prime_k(z_k)$, so the condition
\begin{equation}\label{cond3}
\sum_{j\neq k}\frac{1}{z_k-z_j}=-h'(z_k),\quad 1\leq k\leq n,
\end{equation}
is equivalent to (\ref{cond1}) and (\ref{cond2}). Equation (\ref{cond3})
is the equilibrium condition for
$n$ unit charges at the points $z_k$ in the plane, repelling each
other with the force inversely proportional to the distance, and
in the presence of external field $\overline{h'(z)}$. 
Equations (\ref{cond3}) express the fact that $(z_1,\ldots,z_n)$
is a critical point of the ``master function''
$$\Psi(z_1,\ldots,z_n)=\prod_{(j,k):k<j}(z_k-z_j)\prod_k
e^{h(z_k)}.$$ 

\section{Remarkable identity}\label{22}

From now on we suppose that $h$ is an odd polynomial of degree $3$,
which we write in the form
\begin{equation}\label{h}
h(z)=z^3/3-bz.
\end{equation}
Suppose that all residues of $p^{-2}e^{-2h}$ vanish.
Then the integral
$\int p^{-2}e^{-2h}$ is a meromorphic function in the plane.
Surprisingly, the
integral of some linear combination 
$$\int\left(p^2(-z)e^{-2h(z)}-Cp^{-2}(z)e^{-2h(z)}\right)$$
is not only meromorphic but is
an {\em elementary function}! Here $C$ is a constant depending on $b$ and $p$.
\vspace{.1in}

\noindent
{\bf Conjecture.}
{\em Let $h$ be given by $(\ref{h})$. Let $p$ be a polynomial.
All residues of $p^{-2}e^{-2h}$ vanish if and only if
there exist a constant $C$
and a polynomial $q$
such that
\begin{equation}\label{conj}
\left(p^2(-z)-\frac{C}{p^2(z)}\right)e^{-2h(z)}=
\frac{d}{dz}\left(\frac{q(z)}{p(z)}e^{-2h(z)}\right).
\end{equation}}
\vspace{.1in}

In other words:
$$p^2(z)p^2(-z)-C=q'(z)p(z)-q(z)p'(z)-2q(z)p(z)h'(z).$$

It is known \cite{BB}
that for given $h$ of the form (\ref{h}) there exist
polynomials $p$ of any given degree such that all residues of $p^{-2}e^{-2h}$
vanish.
These polynomials $p$ have simple roots.
We verified the conjecture for $\deg p\leq 4$ by symbolic computation
with Maple.
We don't know whether there is any analog of the Conjecture
for other polynomials $h$.

Substituting $p_n(z)=z^n+az^{n-1}+\ldots $ into (\ref{7}) and using
(\ref{h}), we conclude that 
\begin{equation}\label{P}
P(z)-\lambda=-{h'}^2(z)-h^{\prime\prime}(z)-2nz+2a=-z^4+2z^2b-2(n+1)z-b^2+2a.
\end{equation}
We choose $\lambda=b^2-2a$ so that $P(0)=0$.
Equation (\ref{7}) now becomes
\begin{equation}\label{78}
p^{\prime\prime}_n+2(z^2-b)p^\prime_n-(2nz-2a)p_n=0.
\end{equation}
Coefficients of $p_n$ can be now determined by a linear
recurrence formula.
Putting 
$$p_n(z)=\sum_{j=0}^na_jz^{n-j},\quad a_{-1}=0,\quad a_0=1,\quad a_1=a,$$
we obtain the recurrence
\begin{equation}\label{recur}
ja_j=aa_{j-1}-b(n-j+2)a_{j-2}+\frac{(n-j+2)(n-j+3)}{2}a_{j-3}.
\end{equation}
Coefficients $a_j$ are found from this formula one by one
beginning from $a_1=a$.
Vanishing of the constant term in (\ref{78}) gives a polynomial
equation $Q^*_{n+1}(b,a)=0$ in which we can substitute
$a=(b^2-\lambda)/2$ and write it as
\begin{equation}\label{Q}
Q_{n+1}(b,\lambda)=0.
\end{equation}
Polynomials $p_n,Q^*_{n+1}$ for small $n$ are given in
Appen\-dix~B.

We have $\deg_\lambda Q_{n+1}=n+1,$ \cite{BB}.
For every $b$ and every $\lambda$ sa\-tis\-fy\-ing this equa\-tion,
the dif\-fe\-ren\-tial equation (\ref{1}), with $P$ as in 
(\ref{P}), has a unique so\-lu\-tion $y=p_ne^h$ where $p_n$
is a mo\-nic po\-ly\-no\-mi\-al of degree $n$.
Coef\-fi\-ci\-ents of $p_n$ are poly\-nomials in $b$ and $\lambda$. 

Polynomials $Q_{n+1}$ are fundamental for our subject, but little is known
about them. It seems hard to investigate them
algebraically. In section \ref{77}, we will use analytic tools
to establish some properties of these polynomials,
in particular we will find the terms of top weight and asymptotics
of $\lambda$ as $b\to\infty$.

Functions
$y=p_ne^h$ are eigenfunctions of the operator
\begin{equation}\label{oper}
L_J(y)=y^{\prime\prime}-(z^4-2bz^2+2Jz)y,\quad J=n+1.
\end{equation}
This operator maps the space
$\{ pe^h:\deg p\leq n\}$ of dimension $n+1$ into itself.
For each non-negative integer $n$,
and generic $b$, the operator (\ref{oper}) has $n+1$ eigenfunctions
of the form $p_ne^h$ with
eigenvalues $\lambda$ which are solutions 
of (\ref{Q}). 

We assume without loss of generality
that $Q_{n+1}$ is monic as a polynomial in $\lambda$,
and $p_n$ is a monic polynomial in $z$.
Then the constant $C$ in the Conjecture 
turns out to be
\begin{equation}\label{const}
C(b,\lambda)=
\alpha_n\frac{\partial}{\partial\lambda}Q_{n+1}.
\end{equation}
Symbolic computation for small $n$ shows that $\alpha_n=(-1)^n2^{-2n}$.

\section{Boundary value problem}\label{33}

Eigenfunctions $pe^h$ do not belong to $L^2(\R)$,
but they satisfy the boundary conditions
\begin{equation}\label{bc}
y(te^{\pm\pi i/3})\to 0\quad\mbox{as}\quad t\to\infty.
\end{equation}
With these boundary conditions, the operator (\ref{oper}) is not Hermitian
but PT-symmetric \cite{BB,Shin,Trinh}.

Physicists write the boundary value problem for the operator $L_J$
with the boundary conditions (\ref{bc})
in the equivalent form (\ref{phys}),
which corresponds to the rotation of the independent variable
$i\zeta=z,\; w(\zeta)=y(i\zeta).$ 
We also find this form convenient
in certain arguments,
and will use it in sections \ref{66}, \ref{77}. We keep the notation $y(z)$ for an
eigenfunction of (\ref{oper}), (\ref{bc}), while $w(\zeta)$ stands for an
eigenfunction of (\ref{phys}).

It is known that the boundary value problem
(\ref{oper}), (\ref{bc}) has an infinite sequence
of eigenvalues
tending to infinity \cite{Sibuya}.
Eigenvalues $\lambda$ are solutions of the equation
\begin{equation}\label{specdet}
F_{n+1}(b,\lambda)=0,
\end{equation}
where $F_{n+1}$ is a real entire function on $\C^2$
which is called the {\em spectral determinant} \cite{Shin}.
The set of all solutions of (\ref{specdet}) in $\C^2$
is called the {\em spectral locus} and we denote it by $Z_{n+1}$.
As $F_{n+1}$ is real,
the set of eigenvalues is symmetric with respect to the real line
when $b$ is real.
For each real $b$, all sufficiently large eigenvalues (how large, depends
on $b$) are real \cite{Shin}. 

Equation (\ref{specdet}) is reducible: $F_{n+1}$ is evidently divisible by
$Q_{n+1}$.
On the other hand, equation (\ref{Q}) is irreducible, as follows from
\cite{EG} or \cite{AG}, and 
it defines a smooth algebraic curve in $\C^2$.
This algebraic curve will be denoted by 
$Z_{n+1}^{QES}$.

\section{Theorem 1}\label{44}

Now we discuss a corollary of our conjecture that
we can prove.
Let us fix a simple curve
$\gamma$ in $\C$ parametrized by the real line,
with the properties $\gamma(t)\to\infty,\; \arg\gamma(t)\to\pm\pi/3,\;
t\to\pm\infty$. Then (\ref{conj}) implies
\begin{equation}\label{conj2}
\int_\gamma p^2(z)e^{2h(z)}dz=C\int_{\gamma}p^{-2}(-z)e^{2h(z)}dz.
\end{equation}
To obtain this we replace $z\mapsto-z$ in (\ref{conj}) then integrate
along $\gamma$; the integral in the right hand side of (\ref{conj})
vanishes
because $\Re h(z)\to-\infty$ as $z\to\infty$ on $\gamma$.
Let $\gamma_z$ be a curve consisting of
the piece $\{\gamma(t):-\infty<t\leq 0\}$
followed by a curve from $\gamma(0)$ to $z$. Put
$$g(z)=p(-z)e^{-h(z)}\int_{\gamma_z}p^{-2}(-\zeta)e^{2h(\zeta)}d\zeta.$$
Then $g^-(z):=g(-z)$ satisfies $L_{n+1}(g^-)=\lambda g^-$ so $g$ satisfies
$L_{-n-1}(g)=\lambda g$. To check this, we make a substitution $z\mapsto-z$
in (\ref{oper}).
Moreover, if the integral in the right hand side of (\ref{conj2})
is zero, then $g$ also satisfies the boundary condition (\ref{bc}).
Therefore our conjecture (\ref{conj}), (\ref{const})
has the following corollary:
\vspace{.1in}

\noindent
{\bf Theorem 1.} {\em The points $(b,\lambda)\in Z_{n+1}^{QES}$
where the eigenfunction $y=pe^h$ satisfies
\begin{equation}\label{gamma}
\int_\gamma y^2(z)dz=0
\end{equation}
are either zeros of $C(b,\lambda)$ or points of intersection
of $Z_{n+1}^{QES}$ with $Z_{-n-1}$.}
\vspace{.1in}

We will prove Theorem 1 in the next section.

Equation (\ref{gamma}) is the well-known condition of level crossing,
which we discuss in section \ref{66}.

Thus the Corollary says that the eigenvalues at the points on
$Z_{n+1}^{QES}$ which are singular points of $Z_{n+1}$ are
eigenvalues of two spectral problems, one for $L_{n+1}$ another for
$L_{-n-1}$.

\section{Darboux transform}\label{55}

{\em Proof of Theorem 1.}
Assuming that all residues of $p^{-2}e^{-2h}$ vanish, we will prove 
that the right and left sides of (\ref{conj2}) with $C$ as in
(\ref{const}) have the same
zeros on $Z_{n+1}^{QES}$. Fix the integer $n\geq 0$.
Let $\psi_k=p_ke^h,\quad k=0,\ldots,n$ be all elementary
eigenfunctions of $L_{n+1}$. They are linearly independent,
and they span a space $V$ invariant under $L_{n+1}$.
As $V$ is a subspace of $U=\{ pe^h:\deg p\leq n\}$,
we conclude that $V=U$.
So the Wronski determinant $W=W(\psi_0,\ldots,\psi_n)$ is proportional
to the Wronski determinant 
$$W(e^h,ze^h,\ldots,z^ne^h)=\left(\prod_{k=0}^n k!\right) e^{(n+1)h}.$$
Now let us perform the Darboux transform of $L_{n+1}$ killing these $n+1$
eigenfunctions.
We recall that Darboux transform \cite{Darboux,
Schrodinger,Crum,veselov})
applies
to any operator $-D^2+V$ with eigenfunctions $\phi_0,
\ldots,\phi_n$ and corresponding eigenvalues $\lambda_0,\ldots,\lambda_n$.
The transformed operator is 
$$-D^2+V-2\frac{d^2}{dz^2}\log W(\phi_0,\ldots,\phi_n),$$
and its eigenvalues are those eigenvalues of $-D^2+V$ which are
distinct from $\lambda_0,\ldots,\lambda_n$.
As in our case $2(\log W)^{\prime\prime}=2(n+1)h^{\prime\prime}=4(n+1)z$,
the result of application of the Darboux transform to $L_{n+1}$
and eigenfunctions $\psi_k,\; k=0,\ldots,n$ is $L_{-n-1}$.

If the left hand side of (\ref{conj2}) is zero at some point
$(b,\lambda)\in Z_{n+1}^{QES}$ then $\partial F_{n+1}/\partial\lambda=0$
at this point (see, for example \cite{Trinh}).
As $Z_{n+1}^{QES}$ is smooth, this is possible in exactly two cases:
either $(b,\lambda)$ is a smooth point of $Z_{n+1}$ and
$\partial Q_{n+1}(b,\lambda)\partial\lambda=0$, or $(b,\lambda)$ is a self-intersection
point
of $Z_{n+1}$.

In the second case, $(b,\lambda)$ belongs to the
spectral locus of the Darboux transform $L_{-n-1}$.
This means that the equation
$$L_{-n-1}(y^*)=\lambda y^*$$
with these parameters $(b,\lambda)$
has a solution $y^*$ that tends to $0$ on both ends of $\gamma$.
Then $y_1=y^*(-z)$ tends to $0$ on both ends of $-\gamma$
and satisfies $L_{n+1}(y_1)=\lambda y_1$. So $y_1$ satisfies 
the same differential equation
$L_{n+1}(y)=\lambda y$ as $y$ does, and is linearly independent of $y$.
So $y_1=y\int y^{-2}e^{-2h}$. As this tends to $0$ on both ends of $-\gamma$,
we conclude that
$\int y^{-2}e^{-2h}$ tends to $0$ on both ends of $-\gamma$.
So $y^*(z)=y_1(-z)$
tends to $0$ on both ends of $\gamma$, and this means that the right hand side 
of (\ref{conj2}) is $0$.

The argument is evidently reversible. This proves that the right and
the left hand sides of (\ref{conj2}) have the same zeros
on $Z_{n+1}^{QES}$, that is
(\ref{conj2}) with $C=\alpha_n\partial Q_n/\partial\lambda$,
where $\alpha_n(b,\lambda)\neq 0$ on $Z_{n+1}^{QES}$.

\vspace{.1in}

According to the theorem of Shin \cite{Shin2}, all eigenvalues of $L_J$
for $J\leq 0$ are real. Shin's proof of this uses the ODE-IM correspondence
discovered by Dorey, Dunning and Tateo, \cite{Dorey}.

Combining the Darboux transform used in the prof of Theorem~1 with
the result of Shin \cite{Shin2}, we obtain
\vspace{.1in}

\noindent
{\bf Theorem 2.} {\em For every positive integer $J$, all
non-QES eigenvalues of $L_{J}$ with boundary conditions (\ref{bc})
are real.}
\vspace{.1in}

{\em Proof.} These eigenvalues are also eigenvalues of $L_{-J}$
with boundary conditions (\ref{bc}), and the eigenvalues of $L_{-n-1}$ are
all real by Shin's theorem \cite{Shin2}.
\vspace{.1in}

For $J>1$ there are always some non-real eigenvalues. 

\section{Level crossing}\label{66}

As $Q_{n+1}$ and $F_{n+1}$ are real functions, it is reasonable
to consider real solutions of equations (\ref{specdet}) and (\ref{Q}).
Eigenfunctions $y(z)$ corresponding to these real solutions are
real, while eigenfunctions $w(\zeta)$ (see (\ref{phys})) are PT-symmetric,
that is $w(-\overline{\zeta})=\overline{w(\zeta)}.$
These real solutions $(b,\lambda)$ form curves in $\R^2$ which we call
the {\em real spectral locus} $Z_{n+1}(\R)$ and the
{\em QES real spectral locus}
$Z_{n+1}^{QES}(\R)$,
respectively. 

Now we discuss (\ref{gamma}). First we state a result which
describes $Z_{n+1}^{QES}(\R)$.
\vspace{.1in}

\noindent
{\bf Theorem 3.} {\em For $n\geq 0,$ the spectral locus
$Z_{n+1}^{QES}(\R)$ consists of $[n/2]+1$
disjoint analytic curves
$\Gamma_{n,m},\; 0\leq m\leq[n/2]$
(analytic embeddings of $\R$ to $\R^2$).

For $(b,\lambda)\in\Gamma_{n,m}$, the eigenfunction has $n$ zeros,
$n-2m$ of them real.

If $n$ is odd then $b\to+\infty$ on both ends of each curve
$\Gamma_{n,m}$. If $n$ is even then the
same holds for $0\leq m<n/2$, but on the ends of
$\Gamma_{n,n/2}$ we have $b\to\pm\infty$. 

If $(b,\lambda)\in \Gamma_{n,m},\; (b,\mu)\in\Gamma_{n,m+1}$
and $b$ is sufficiently large, then $\mu>\lambda$.
} 
\vspace{.1in}
 
The proof of this theorem 
can be found in \cite{EG7}. It follows the method of \cite{EG2}
where similar results were established for real spectral loci
of other families of cubic and quartic potentials.
The method is based on singular perturbation
and Nevanlinna parametrization of the spectral locus.

Computer generated pictures of $Z_{n+1}(\R)$ show an interesting
phe\-no\-me\-non:
when $n$ is even, the curve $\Gamma_{n,n/2}$ crosses the non-QES
part of the spec\-tral locus \cite[Fig. 1]{BB}. 
We will prove that infi\-ni\-tely many such cros\-sings exist for
even $n$ and negative
$b$.

We say that a level crossing occurs at a point $(b,\lambda)$
of the spectral locus
if $\partial F_{n+1}/\partial\lambda=0$ at this point. 
If $y$ is the eigenfunction corresponding to a point $(b,\lambda)$,
then
the level crossing occurs if and only if (\ref{gamma}) is satisfied 
\cite[II.7]{Simon}, \cite[Thm. 8]{Trinh}.
There are two types of level crossing points:
\vspace{.1in}

\noindent
a) Critical points of the function $\lambda$ at non-singular
points of $Z_{n+1}$. 

If such a critical point $(b_0,\lambda_0)$
is simple and belongs to $Z_{n+1}(\R)$ then
the two eigenvalues that meet at this point are both real for
$b$ on one side of $b_0$ and complex conjugate on the other side.
\vspace{.1in}

\noindent
b) Singular points of $Z_{n+1}$.

If two eigenvalues collide at a
simple self-intersection point of
$Z_{n+1}(\R)$ with two distinct non-vertical tangents, then these eigenvalues
both remain real in a neighborhood of $b_0$.
Operator $L_{n+1}$ with $b=b_0$ contains a Jordan cell in this case.
\vspace{.1in}

We recall that $Z_{n+1}^{QES}$ is a smooth curve.
Thus the crossing points on $Z_{n+1}^{QES}$ where only QES eigenvalues collide
are all of type a), and they satisfy 
$$Q_{n+1}(b,\lambda)=0,\quad \frac{\partial}{\partial\lambda}Q_{n+1}(b,\lambda)=0.$$
For each $n$, there are finitely many such points on $Z_{n+1}^{QES}$. 

We will show that there are always infinitely many crossing points 
of type b) where QES eigenvalues collide with non-QES eigenvalues.
So the curve defined by (\ref{specdet}) is not smooth: it has
infinitely many self-intersections.

We don't know whether more complicated singularities than a) and b)
exist; numerical experiments only show singularities of
types a) and b). 
\vspace{.1in}

\noindent
{\bf Proposition 1.} {\em Function
$$\Phi_n(b,\lambda)=\int_\gamma y^2(z)dz,\quad Z_{n+1}^{QES}\to\C,$$
where $y$ is the eigenfunction corresponding to $(b,\lambda)$,
has infinitely many zeros $(b_k,\lambda_k),\; b_k\to\infty.$
When $n$ is even, $\Phi_n$ has infinitely many zeros with negative $b_k$
and real $\lambda_k$.}\vspace{.1in}

{\em Proof.} We have 
$$\Phi_n(b)=\int_\gamma p_n^2(z)e^{2h(z)}dz.$$
We remind that coefficients of $p_n$ and $h$ are algebraic functions
of $b$. 
When $n=0$, we can take $p_0=1$, and then 
$$\Phi_0(b)=\int_\gamma e^{(2/3)z^3-2bz}dz=2^{2/3}i\pi\Ai(2^{2/3}b),$$
where $\Ai$ is the Airy function \cite{AS}. Airy function
is a real entire function
of order $3/2$ with infinitely many negative simple zeros. 

To generalize this to other values of $n$,
we express $\Phi_n$ as a linear combination of
$\Phi_0$ and $\Phi_0^\prime$ with coefficients depending on $b$ algebraically.
Differentiating
$\Phi_0(b)$
with respect to $b$, we obtain
$$\int_\gamma z^k e^{(2/3)z^3-2bz}dz=(-2)^{-k}\Phi_0^{(k)}(b),$$
and thus
$$\Phi_n(b)=p_n^2(-D/2)\Phi_0(b),$$
where $D=d/d b$. Now
all $\Phi_0^{(k)}$ are linear combinations of $\Phi_0$ and $\Phi_0^\prime$
with polynomial
coefficients because $\Ai$ satisfies the differential equation
$\Ai^{\prime\prime}(s)=s\Ai(s).$ So $\Phi_n$ is of the form
\begin{equation}\label{ode}
\Phi_n(b)=A_n(b)\Phi_0(b)+B_n(b)\Phi_0^\prime(b),
\end{equation}
where $A_n$ and $B_n$ are algebraic functions.

We claim that every linear combination $\phi$ of $\Phi_0$ and $\Phi_0^\prime$
with algebraic coefficients has infinitely many zeros. We prove this claim
by contradiction. Suppose that such a linear combination 
\begin{equation}
\label{lc}
\phi=a_0\Phi_0+a_1\Phi_0^\prime
\end{equation}
has finitely
many zeros. Let $F$ be a compact Riemann surface
spread on the Riemann sphere on which $a_0$ and $a_1$ are meromorphic.
Then $\phi$ is meromorphic on $F\backslash E$, where
$E$ is the finite set of points of $F$ lying over $\infty$.
At the points of $E$, $\phi$ has
isolated essential
singularities. 
As $\phi$ has finitely many zeros and poles on $F\backslash E$, we conclude
that $\phi'/\phi$ is meromorphic on $F\backslash E$.
The growth estimate $\log|\phi(b)|\leq O(|b|^{3/2})$, $b\to\infty$,
implies that the points of $E$ are removable singularities
of $\phi'/\phi$. Thus $\phi$ is the exponent of an Abelian integral.
Now consider (\ref{lc}) as a linear differential
equation of first order with respect to $\Phi_0$,
whose coefficients belong to the minimal field
$K$ that contains $\C(b)$, is algebraically
closed, and contains a primitive of every element, and the exponent of a 
primitive of every element.
As every first order linear differential equation can be solved by integration
we conclude that $\Phi_0\in K$ which implies that $\Ai\in K$.
But this is not so by a well-known classical theorem of Picard and
Vessiot, 
\cite[Theorem 6.6]{Kaplansky}.
This proves our claim.

When $n$ is even, according to Theorem~3, we have a real analytic branch
$\lambda(b)$ defined for all real $b$ with sufficiently large
absolute value. The graph of this branch is a part of $\Gamma_{n,n/2}$.
Using this branch we rewrite the equation $\Phi_n(b)=0$
as 
$$\Phi_0^\prime(b)/\Phi_0(b)=A(b),$$
where $A$ is a real branch of an algebraic function on $(-\infty,B)$
with some $B\in\R$.
This last equation has infinitely many negative solutions
because $\Phi_0$ has infinitely many negative zeros and they are
interlaced with zeros of $\Phi_0^\prime$.
This completes the proof of the proposition.
\vspace{.1in}

Using the asymptotics of the zeros of Airy's function \cite{AS}
we obtain that the crossing points
satisfy $b_k\sim-((3/4)\pi k)^{2/3}$, $k\to\infty$.

\section{Asymptotics as $b\to+\infty$}\label{77}

Now we study asymptotics of the eigenvalues $\lambda$
as $b\to+\infty$ and make conclusions about polynomials $Q_{n+1}$.
Our main result here is the explicit formula (\ref{quasi}) for the
top quasi-homogeneous part of $Q_{n+1}^*$.

First we obtain a preliminary estimate of solutions $\lambda(b)$
of equation (\ref{Q}) for large $b$:
\begin{equation}\label{lam}
\lambda(b)\sim b^2+O(\sqrt{b}),\quad b\to\infty.
\end{equation}
To prove this, consider the recurrence (\ref{recur}). For a monomial 
$a^mb^k$ we define the weight as $m+2k$.
Then (\ref{recur}) implies that 
$$j!a_j=a^j+\sum_{m=1}^{[j/2]}c_{m,j}b^ma^{j-2m}+\mbox{terms of lower weight}.$$
Vanishing of the constant term in (\ref{78}) gives
$$Q_{n+1}^*(a,b)=aa_n+ba_{n-1}+2a_{n-2}=0,$$
so $Q_{n+1}^*$ is a sum of a quasi-homogeneous polynomial in $a$ and $b$
of weight $2(n+1)$ and a polynomial of lower weight.
This means that $a=O(\sqrt{b})$ and $\lambda(b)=b^2-2a$
satisfies (\ref{lam}).

To obtain more precise asymptotics we use singular perturbation
arguments from \cite{EG2}, which we state in Appendix A for the reader's
convenience.

Suppose that $b$ is real and
$b\to+\infty$.
In the equation (\ref{phys}) we set 
$$\zeta=\epsilon u-i\epsilon^{-2},\quad b=\epsilon^{-4},\quad
 W(u)=w(\epsilon u-i\epsilon^{-2}).$$
The result is
\begin{equation}\label{resc}
W^{\prime\prime}+(\epsilon^6u^4-4i\epsilon^3u^3-4u^2-2iJ\epsilon^3u)W
+(2J+\epsilon^2\lambda-\epsilon^{-6})W=0,
\end{equation}
or
\begin{equation}\label{qeslam}
-W^{\prime\prime}-\left(u^2(b^{-3/4}u-2i)^2-2iJb^{-3/4}u\right)W=
(2J+b^{-1/2}\lambda-b^{3/2})W.
\end{equation}
When $\epsilon\to 0$, we obtain the limit eigenvalue problem
\begin{equation}
\label{harm}
-W^{\prime\prime}+4u^2W=\mu W,
\end{equation}
which is a harmonic oscillator with eigenvalues $\mu_k=2(2k+1),\;
k=0,1,2,\ldots$. By a general result from \cite{EG2} (see Appendix),
(\ref{qeslam}) implies that
for each $k$, there
must be a unique eigenvalue $\lambda_k(b)$ which satisfies
\begin{equation}\label{k}
\lambda_k=b^2+(\mu_k-2J+o(1))\sqrt{b}.
\end{equation}
Moreover, for each compact set $K$
in the $\lambda$-plane there exists
$b_0>0$ such that for $b>b_0$ there are no other eigenvalues $\lambda(b)\in K$,
except those satisfying (\ref{k}).

We conclude from (\ref{lam}) that QES eigenvalues must
satisfy (\ref{k}). 
That is for each QES eigenvalue $\lambda$ there exists $k$
such that (\ref{k}) holds.
Now we have to find out what are the values of $k$ for the QES eigenvalues.

To do this, we consider zeros of eigenfunctions.
We know that $k$-th eigenfunction of (\ref{harm}) has $[k/2]$
zeros in the right half-plane, the same number of zeros in the
left half-plane, and one zero on $i\R$ if $k$ is odd.
(In fact all these last zeros belong to the real line but this is irrelevant
for our argument.)
So for every $m=0,1,\ldots$
there are two
eigenfunctions of the harmonic oscillator
(with $k=2m$ and $k=2m+1$) which have $m$
zeros in the right half-plane, and one of them ($k=2m+1$) has
a zero on $i\R$.

Theorem 3 implies that for
each given $n$ and for each $m\leq [n/2]$
and $b$ sufficiently large positive, there is exactly one curve $\Gamma_{n,m}$,
such that the corresponding eigenfunctions  have
$m$ zeros in the right half-plane\footnote{Remember
that we are working here with eigenfunctions $w(\zeta)=y(i\zeta)$, where
$y$ is an eigenfunction from Theorem 3.}.
We refer to \cite{EG2} for the argument showing that the zeros
of eigenfunctions $w$ in the right half-plane do not escape to
infinity as $b\to+\infty$. Zeros of $w$ on $i\R$ 
{\em do} escape to infinity, except possibly one of them. 
Thus the branches of QES eigenvalues must be $\lambda_0,\ldots,\lambda_n$
satisfying (\ref{k}).

Putting $\lambda_k=b^2-2a(k),$ 
and $J=n+1$ in (\ref{k})
we obtain
$$a(k)\sim \sqrt{b}(n-2k),\quad 0\leq k\leq n.$$
We conclude that the top weight term of the polynomial $Q_{n+1}^*$
is
\begin{equation}
\label{quasi}
\prod_{k=0}^n\left( a-(n-2k)\sqrt{b}\right)
=\left\{
\begin{array}{ll}(a^2-b)(a^2-3b)\ldots(a^2-nb),& n\;\mbox{is odd},\\
a(a^2-2b)\ldots(a^2-nb),& n\;\mbox{is even}.
\end{array}
\right.
\end{equation}
This implies that the degree of the discriminant of $Q_{n+1}^*$
is $n(n+1)/2$, and the genus of the QES spectral locus is 
$n(n-2)/4$ when $n$ is even and $(n-1)^2/4$ when $n$ is odd.

\section{Asymptotics as $b\to-\infty$}\label{88}

When $b\to-\infty$, our operator (\ref{phys}) also 
degenerates to a harmonic oscillator. However none of the QES eigenvalues
of (\ref{phys}) tend to the eigenvalues of this harmonic oscillator
as $b\to-\infty$.
To study this limit, we set $z=\epsilon u,\; b=-\epsilon^{-4}$
and $W(u)=w(\epsilon u)$ in (\ref{phys}).
The result is
\begin{equation}\label{25}
W^{\prime\prime}+\left(\epsilon^6u^4-2u^2+2iJ\epsilon^3u+
\epsilon^2\lambda\right)W=0.
\end{equation}
As $\epsilon\to 0$, this tends to the harmonic oscillator
$$-W^{\prime\prime}+2u^2W=\mu W,$$
whose eigenvalues are $\mu_k=\sqrt{2}(2k+1)$, $k=0,1,2\ldots.$
So by the results in \cite{EG2} (see Appendix A),
for every $k$ and for $b<-b_k$,
there is an eigenvalue $\lambda_k(b)$ which satisfies
$$\lim_{\epsilon\to 0}\epsilon^2\lambda_k(b)=\sqrt{2}(2k+1),$$
or $\lambda_k(b)\sim \sqrt{-b}.$ Comparison with (\ref{lam}) shows
that these eigenvalues $\lambda_k$ cannot come from the QES spectrum.
\vspace{.1in}

We thank Per Alexandersson for making Fig.~1, and for help with computations
which led us to the discovery of (\ref{conj}), (\ref{const}),
Stefan Boettcher for 
sending us pictures of $Z_J(\R)$ which inspired this work,
Evgenii Mukhin and Alexandre Varchenko
for useful discussions, and 
Vladimir Marchenko for his insightful suggestion
to look at the Darboux transform.
\newpage

\appendix
{\bf \Large Appendix A. Singular perturbation of polynomial potentials}
\vspace{.2in}

Here we state the main singular perturbation result of \cite{EG2}
and verify that the eigenvalue problems (\ref{resc}) and (\ref{25}) satisfy
all conditions that imply continuity of the discrete spectrum at $\epsilon=0$.

Consider the eigenvalue problem
\begin{equation}\label{a1}
-y^{\prime\prime}+P_\epsilon(z,b)y=\lambda y,
\quad y(z)\to 0,\quad z\in R_1\cup R_2.
\end{equation}
Here $z$ is the independent variable, $P$ is a polynomial in $z$ whose
coefficients depend on parameters $\epsilon>0$ and $b\in\C$,
dependence on $b$ is holomorphic,  and $R_1,R_2$ are two rays
in the complex plane defined by $R_k=\{ te^{i\theta_k}\in\C_z: t>0\}$,
$k=1,2$.

Suppose that 
$$P_\epsilon(z,b)=\sum_{j=0}^d a_j(b,\epsilon)z^j,$$
where 
$a_d(\epsilon)>0$ does not
depend on $b$,  
$P_0(z,b)=a_m(b,0)z^m+\ldots$, where $m<d$,
and 
the dots stand for the terms of smaller degree in $z$.

Let 
$$P^*_\epsilon(z,b)=\sum_{j=m}^d a_j(b,\epsilon)z^j.$$

For every polynomial potential $P(z)=a_nz^n+\ldots$ of degree $n$,
the {\em separation rays}
are defined by
$$\{ z\in \C:a_nz^{n+2}<0\}.$$
{\em Turning points} are just zeros of the potential $P$
in the complex plane.\footnote{This terminology is somewhat unusual
but convenient here. In the standard terminology turning points
are zeros of $P-\lambda$.}
{\em Vertical line} at a point $z$ is the
line defined by $P(z)dz^2<0$. If $P$ depends on parameters, 
then the separation rays, turning points and the vertical line field
depend on the same parameters.

We assume that there exists $\delta>0$ and $\epsilon_0>0$ and
a compact $K\subset\C_b$, such that for all
$\epsilon\in(0,\epsilon_0)$ and for all $b\in K$ and $k\in\{1,2\}$ the following
conditions are satisfied:
\vspace{.1in}

(i) $|\arg z-\theta_k|\geq\delta$ for all turning points
$z\in\C\backslash\{0\}$ of $P^*_\epsilon$,
\vspace{.1in}

(ii) For every point $z\in R_k$, the smallest angle between
$R_k$ and the vertical line with respect to $P^*_\epsilon$ at this point
is at least $\delta$.
\vspace{.1in}

(iii) 
$R_k$ are not separation rays, for $P_\epsilon,\; \epsilon>0$
or $P_0$. 
\vspace{.1in}

(iv) All coefficients $a_j(b,\epsilon)$ are bounded from above
and $|a_m(b,\epsilon)|$ is bounded from below.
\vspace{.1in}

\noindent
{\bf Theorem A.} {\em  If the conditions (i)--(iv) are
satisfied, then the spectral determinant $F_\epsilon$
of the eigenvalue problem (\ref{a1}) converges as $\epsilon\to 0$
to the spectral determinant of (\ref{a1}) with $\epsilon=0$:
$$F_\epsilon\to F_0,\quad\epsilon\to 0,$$
uniformly for $(b,\lambda)\in K\times K_1$, for every compact $K_1\subset\C_z$.}
\bigskip
\begin{center}
\epsfxsize=2.5in%
\centerline{\epsffile{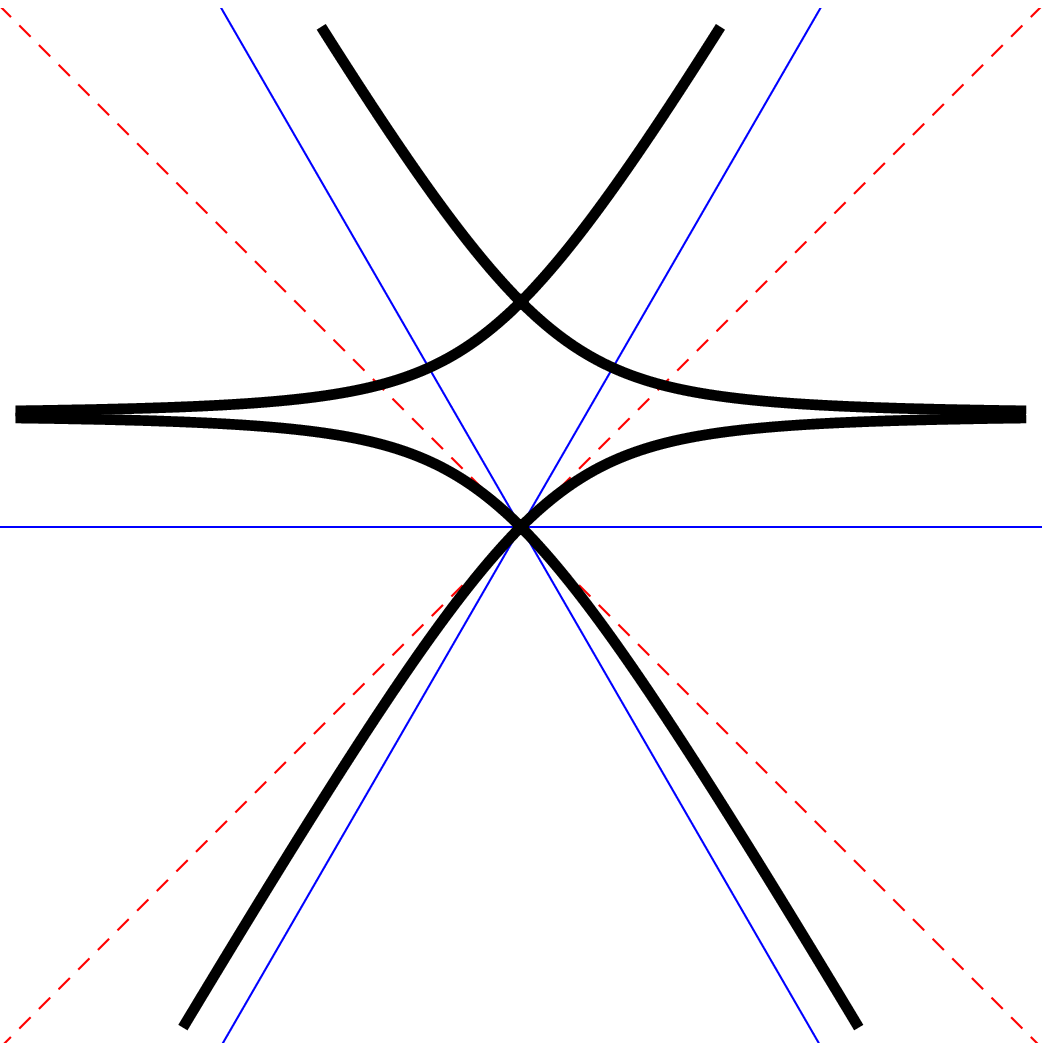}}
\nopagebreak\medskip
\noindent Fig 1. Stokes complex of $P^*_\epsilon$.
\end{center}
\bigskip

Now we verify that the family of potentials in (\ref{resc})
satisfies all conditions (i)-(iv) with $d=4,m=2$.
We have 
$$P^*_\epsilon(z)=-\epsilon^6z^4+4i\epsilon^3z^3+4z^2,$$
$P_0^*(z)=4z^2$.
The turning points are $0$ and $2i\epsilon^{-3}$.
The separation rays are $\arg z\in\{ 0,\pi,\pm\pi/3,\pm 2\pi/3\}$
for $P^*_\epsilon, \epsilon>0$ (shown in thin solid lines in Fig.~1),
and $\arg z\in\{ \pm\pi/4,\pm 3\pi/4\}$ for $P_0$
(dashed lines in Fig.~1).
The normalization rays are $\arg z\in\{ -\pi/2\pm\pi/3\}$.
The bold lines in Fig.~1 represent the Stokes complex,
that is the integral curves of the vertical
direction field $P^*_\epsilon(z)dz^2<0$
that are adjacent to the turning points.

Thus conditions (i),(iii) and (iv) evidently hold.
It remains to verity (ii).

To do this we parametrize $R_1$ as $z=te^{-i\pi/6}:\; t>0$
and find the direction of the line field $\arg dz$ at $z$
by inserting this parametrization to
$\arg(P^*_\epsilon(z)dz^2)=\pi$.
We obtain
$$\arg P^*_\epsilon(z)\in(-\pi/2,\pi/3),
\quad\pm\arg dz^2\in (2\pi/3,4\pi/3),$$
so the angle between $dz$ and $R_1$ is at least $\pi/6$.
Verification for $R_2$ is similar.

We leave to the reader to verify that conditions of Theorem~A are satisfied
for (\ref{25}).
\vspace{.2in}

{\bf \Large Appendix B. Explicit expressions}
\vspace{.2in}

We remind that $\lambda=b^2-2a$ where $p_n(z)=z^n+az^{n-1}+\ldots$.
Since $Q_{n+1}(b,\lambda)$ and $Q^*_{n+1}(b,a)$ are normalized so that
they are monic polynomials in $\lambda$ and $a$, respectively,
we have $Q^*(b,a)=(-1)^{n+1}Q_{n+1}(b,b^2-2a)/2^{n+1}$.
We use the notation $C^*(b,a)=C(b,b^2-2a)$, where $C$ is the
constant from (\ref{conj}). Then (\ref{const}) can be rewritten as
\begin{equation}\label{consta}
C^*(b,a)=
\alpha^*_n\frac{\partial}{\partial a}Q^*_{n+1},
\end{equation}
where $\alpha^*_n=(-1)^n 2^n \alpha_n=2^{-n}$.

Here are results of symbolic computations with Maple.

\noindent
For $n=1$:
$$p_1(z)=z+a,$$
$$Q^*_2(b,a)=a^2-b,$$
$$C^*(b,a)=a.$$
For $n=2$:
$$p_2(z)=z^2+az+\left(\frac{a^2}{2}-b\right),$$
$$Q_3^*(b,a)=a^3-4ab+2,$$
$$C^*(b,a)=\frac{3}{4}a^2-b=\frac{1}{4}\frac{\partial Q^*_2}{\partial a}.$$
For $n=3$:
$$p_2(z)=z^3+az^2+\left(\frac{1}{2}a^2-\frac{3}{2}b\right)z-\frac{7}{6}ab
+\frac{1}{6}a^3+1,$$
$$Q_4^*(b,a)=a^4-10a^2b+12a+9b^2,$$
$$C^*(b,a)=\frac{1}{2}a^3-\frac{5}{2}ab+\frac{3}{2}=\frac{1}{8}\frac{\partial
Q^*_4}{\partial a}.$$
For $n=4$:
$$p_4(z)=z^4-az^3+\left(\frac{1}{2}a^2-2b\right)z^2-\left(2+\frac{1}{6}a^3-
\frac{5}{2}ab\right)z-\frac{2}{3}a^2b+b^2+\frac{5}{4}a+\frac{1}{24}a^4,$$
$$Q_5^*(b,a)=42a^2-96b-20a^3b+64ab^2+a^5,$$
$$C^*(b,a)=\frac{21}{4}a+\frac{5}{16}a^4-\frac{15}{4}a^2b+4b^2=\frac{1}{16}
\frac{\partial Q^*_5}{\partial{a}}.$$

{\em Department of mathematics

Purdue University

West Lafayette IN 47907

eremenko@math.purdue.edu

agabriel@math.purdue.edu}
\end{document}